\documentclass[twocolumn,twoside,slac,a4paper]{revtex4}
\usepackage{graphicx}
\usepackage{fancyhdr}
\usepackage[dvips]{epsfig}
\begin{document}

\title{AMANDA - first running experiment to use GRID in production}

%

\author{T. Harenberg, K.-H. Becker, W. Rhode, C. Schmitt}
\affiliation{University of Wuppertal, Fachbereich Physik, 42097 Wuppertal, Germany}
%

\begin{abstract}
The Grid technologies are in ongoing development. Using current 
Grid toolkits like
the Globus toolkit \cite{wwwglobus} gives one the possibility to build
up virtual organizations as defined in~\cite{anatomy}.
Although these tookits
are still under development and do not feature all functionality,
they can already now be used to set up an efficient computing
environment for physics collaborations with only moderate work. We
discuss in this paper the use of such a computing structure in two running
experiments - the AMANDA (AMANDA = Antarctic muon and neutrino
detector array) neutrino telescope and the D\O \  experiment at Tevatron,
Fermilab. One of the main features of our approach is to avoid reprogramming of
the existing software which is based on
several programming languages (FORTRAN, C/C++, JAVA). This was
realized with 
software layers around the collaboration software taking care
about in- and output, user notification, tracking of running jobs,
etc. 
A further important aspect is the resolution of library
dependencies, which 
occur when a user runs self-compiled jobs on machines where these libraries
are not installed. These dependencies are also resolved with these
layers.

\end{abstract}

\maketitle

\thispagestyle{fancy}







\section{Introduction}
\label{intro}

AMANDA is a running neutrino telescope 
situated at the south pole. Its collaboration
members are spread throughout the world, mainly in North America and
Europe. Our aim was to help this collaboration to create a
user-friendly, unified access to the standard software repository using
existing GRID toolkits. Furthermore, the spreaded computing power of
the participating institutes should be not united, but the access to
foreign resources should be unified to give any single physicists
within the collaboration the possibility to have appropriate computing
power available when needed.

Like in other experiments, the standard simulation software within
AMANDA has a grown
structure and consists of many parts written by several people and in
several programming languages (FORTRAN, C/C++, JAVA).
 To use the GRID software structures,
some reprogramming would be needed, which is difficult in running
experiments. We show how we solved this with an approach which
required no reprogramming of the existing software.

At D\O \ the situation is different: the experiment 
has a data access system 
SAM\footnote{SAM = Sequential data Access via Meta-data}\cite{wwwsam} 
which 
implements some of the basic GRID ideas. Our emphasis here was to
show that for some analyses our system can be used as a queuing
system to prove that GRID- and non-GRID-parts work smoothly together.

In this paper we start by briefly summerizing the basic idea of the
GRID and its different layers and of the different protocols used in such an
environment. After that, we describe the GRID system which has been
built up at our institute and extended to further collaboration
members.

Afterwards, the parts of the collaboration software are identified which
have to be rewritten to use this GRID infrastructure. To minimize the
changes 
we use a different  approach at
AMANDA, which is being presented. And we show that also within D\O \  
our system can be
used. Finally, we present the graphical user interface
(GUI) which has been developed to give the physicists an easy access to
the GRID.

\section{An Introduction to GRID}

This basic idea of ``The Grid'' is defined in the Book
{\it The Grid: Blueprint for a New Computing Infrastructure} by
I.~Foster and C.~Kesselman\cite{foster} as: {\it Infrastructure that
enables the integrated,
collaborative use of high-end computers, networks, databases, and
scientific instruments owned and managed by multiple organizations.}
One may compare the GRID to the electrical power grid, where the the
power plants and the consumers are connected via a large network. The
single consumer is not interested where his energy comes from. This
is one of the main topics of building up a grid: the user should gain
access to all kinds of computer resources (CPU, storage, data, \dots)
without having to care about the underlying access procedures. The
GRID provides him with a {\bf uniform} access mechanisms to all kinds
of resources. 

\begin{figure}
  \begin{center}
    \mbox{
            \epsfig{file=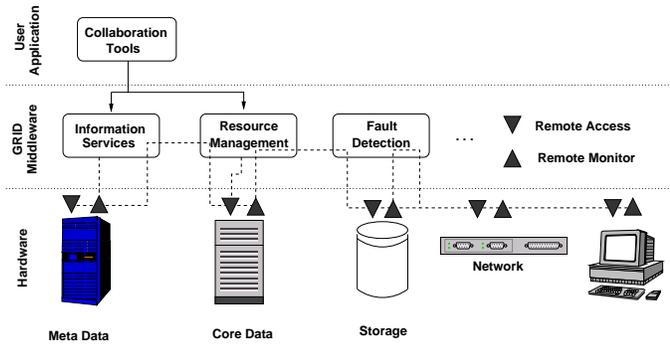,height=4.5cm}
         }
  \end{center}
  \caption[The Role of GRID Services (aka Middleware) and Tools]
  {The Role of GRID Services (aka Middleware) and Tools}
  \label{layer}
\end{figure}

To achieve this, the Globus toolkit introduces a software structure,
the so called ``middleware'', which is a software layer in between the
user application and all kinds of 
resources, as shown in Figure~\ref{layer}. 
Our work is based on 
the Globus toolkit~\cite{wwwglobus}, which is a common
toolkit for GRID developments. 

The ``middleware'' introduces some
protocols which are used by the participating computers to
communicate. These protocols deal with the authentication of users,
the data transfer, and the information interchange to allocate free
resources. The following table~\ref{protocols} gives an overview of
these protocols
and some of its non-GRID equivalents. Note that this table is not
complete and due to the ongoing developments things may change
quickly. The last column shows some commands which are implemented in
the Globus toolkit to access these protocols from the shell. Within
our project all protocols are used with the exception of the Heartbeat
Monitor.

\begin{table}[h]
 \begin{center}
  \begin{tabular}{|p{2cm}|p{2cm}|p{2cm}|p{2cm}|p{2cm}|p{2cm}|} \hline
   Task & Standard Protocol & Grid equivalent & command-line tool(s) \\
   \hline
   \hline
   Access to machines & telnet, rlogin, rsh, ssh, \dots & GRID
Security Infrastructure~\cite{SECURITY} & globusrun, globus-job-run, \dots  \\
   \hline
   Data transfer & ftp, scp, \dots & Globus Access to Secondary
Storage (GASS)~\cite{GASS}, GridFTP~\cite{gridftp} & globus-url-copy \\
   \hline
   Resource finding & N/A (external software like OpenLDAP) & Metadata
Directory Service (MDS)~\cite{MDS} & ldapsearch \\
   \hline
   Computer monitoring & N/A (external watchdog software) & Heartbeat
Monitor (HBM)~\cite{HBM} & N/A\\
   \hline
 \end{tabular}
\end{center}
\caption[Overview over Grid protocols]
{\sl Overview over Grid protocols}
 \label{protocols}
\end{table}

\section{A GRID for AMANDA and for D\O}
\label{chapter_grid_for_amanda}


Although current GRID toolkits like the used Globus Toolkit 2 have
all basic features needed to build up a GRID included, not all ideas
could be implemented without major programming Globus Toolkit
itself.

For AMANDA, we focussed the following:

\begin{itemize}
\item the user should have access to a central software repository,
where the standard offline software is ``ready to use'',
\item the user should be able to use the batch queuing systems in
its own institute and in other institutes participating in AMANDA without
having to care about the local infrastructure, the access policies and
the network infrastructure (Firewalls),
\item the in- and output files of the software should be transparent
to the user, which means that for him it makes no different where the
code runs,
\item a list of running and finished jobs should be available to the
user, this is a job not covered by the Globus Toolkit yet,
\item for mass production the generated data should be available at the
centralized data storage,
\item besides standard software, ``own code'' should also be possible
to run within the GRID environment, file transfer should be provided
and
\item the software should care that own user code should run on remote
sides even if the binary has been dynamically linked to libraries
which may be not installed at the remote side.
\end{itemize}

These features has been postponed for later versions:

\begin{itemize}
\item the system does not search an appropriate batch system itself,
as the needs of the software cannot be guessed from the binary. This
is especially true for software programmed by the user,
\item ports to other operating systems
\end{itemize}

As mentioned in the introduction, the software within AMANDA shows a
grown structure with a variety of programs written in many different
programming languages (C, JAVA, FORTRAN, \dots). Although a JAVA port
of the Globus toolkit exists, enabling FORTRAN code to use the Globus
protocols seems to be unfeasable.

Therefore, some code has been written around the standard software which
takes care of the in- and output using the GRID protocols. The
standard software runs in a kind of sandbox, where all the necessary
libraries and files are provided, and after the end of a job, 
the output is transfered back. For this reason, we create a
software server, serving the standard software together with the code
around as a bundle. Every time a user requests a run, this software is
transfered to the executing node by the queuing system and then
executed in a temporary file space. After job termination, all files are
cleared up. This is explained in more detail in the following
chapter~\ref{kapwugrid}. And we show that we also can use our system
without the Grid components, therefore we present an example where we
used it as a queuing system for D\O.

\subsection{Thoughts towards a GRID in Wuppertal}
\label{kapwugrid}

This chapter first introduces to the situation at the Physics
Department in Wuppertal to enhance what a GRID should achieve. After
that, the installation of a GRID system in Wuppertal is explained and
the Gridnavigator program, which has been developed here, is
presented.

The groups in Wuppertal involved in AMANDA and D\O\  experiments have a
well equipped computer infrastructure, but no centralized
INTEL-CPU based computing cluster. The aim of our work
was to make the CPU power usable which is available in desktop PCs,
which are of Pentium-III 1 GHz class. The development of our software
was done on several PCs running different flavors of Linux (SuSE Linux
Versions from 7.3 to 8.1 and RedHat 7.x). Porting to other platforms
as for example DEC alpha has been postponed.


To explain the difference between a GRID system and a conventional
batch queuing system we first take a look how a typical conventional approach
to set up a queuing system looks like:

The machines are connected via a (fast) local area network (LAN). 
One machine acts as a central server which holds the disks and a
central account service. This disk space is - together
with the account information - exported to the cluster machines using
protocols like the Network File System (NFS) and the Network
Information Service (NIS).

\begin{figure}
  \begin{center}
    \mbox{
            \epsfig{file=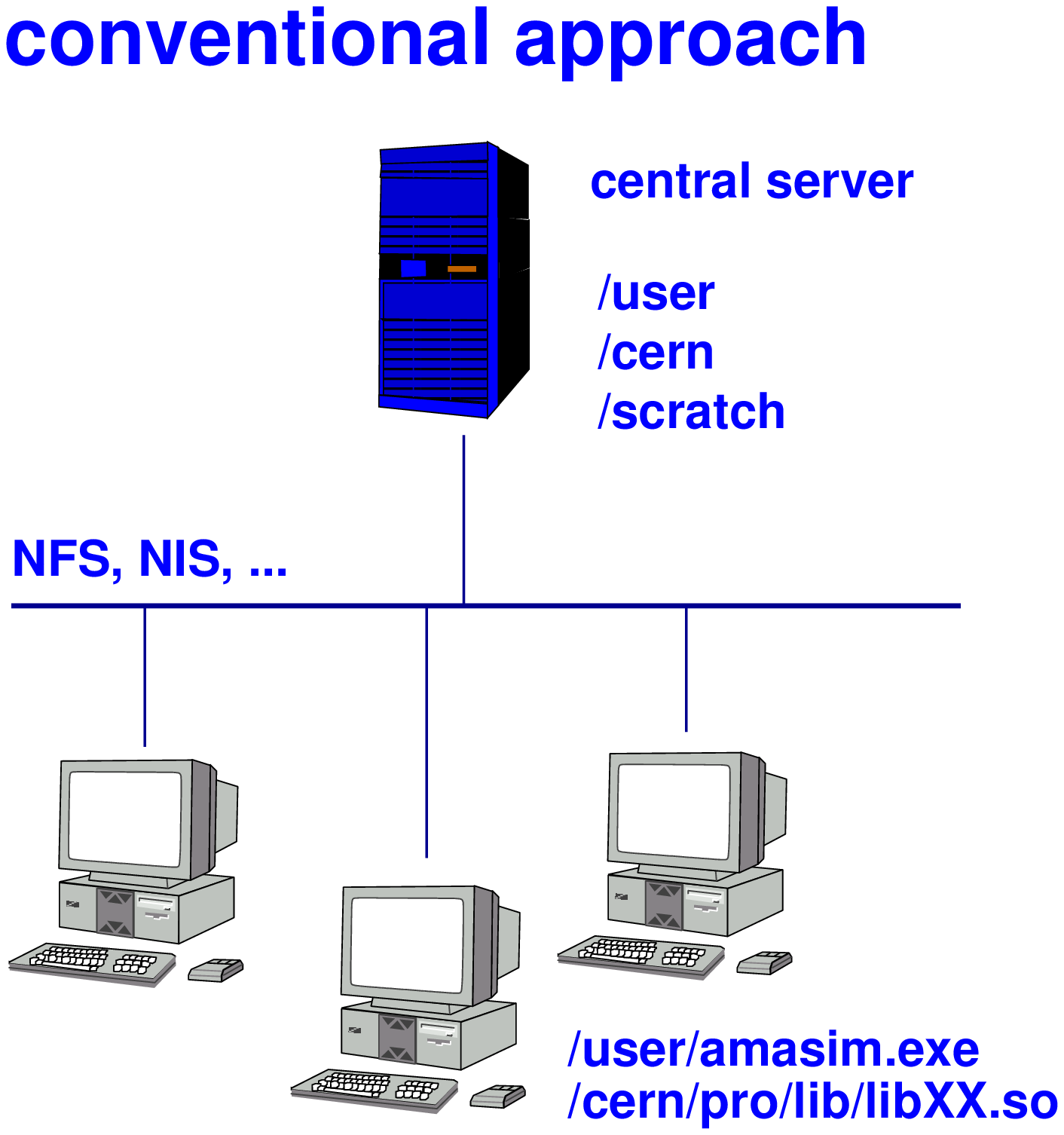,height=7.5cm}
         }
  \end{center}
  \caption[A conventional approach to set up a batch queuing system]
  {A conventional approach to set up a batch queuing system}
  \label{sandbox}
\end{figure}

The computer program is executed at one
of the cluster machines, but doing so every file (executable, library,
program data) is transfered at the moment it is needed via the network
to the executing machine. Furthermore, every file created or changed
by the software has to be transfered ``online'' back to the
server, as illustrated by fig.~\ref{sandbox}. This is normally not a
problem in fast LANs, but on slower and less reliable wide area
networks (WANs), this structure may result in slow job execution, high
network load and
may completely stop when the network is down for even a short amount
of time. 

\begin{figure}
  \begin{center}
    \mbox{
            \epsfig{file=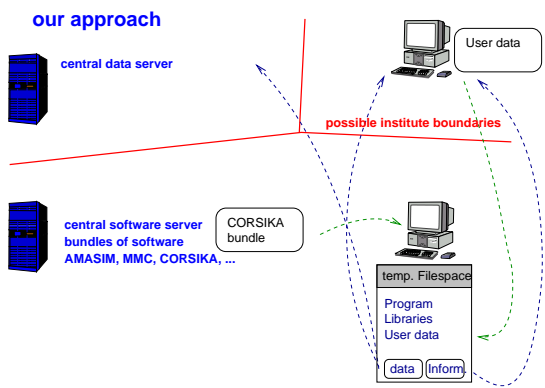,height=5cm}
         }
  \end{center}
  \caption[The Wuppertal approach to build up a GRID]
  {The Wuppertal approach to build up a GRID}
  \label{sandbox_grid}
\end{figure}

Our approach is different and shown in fig.~\ref{sandbox_grid}.

We use the GRID protocols to gain access to the machines and to
transfer data. We bundle all the software together with the sandbox
software, as introduced in chapter~\ref{chapter_grid_for_amanda}. This
software is stored on a central server which can be unique for every
institute. A
central data server is set up - this is a machine with a large RAID
disk array to store the data in mass production. After the machine
which should run the software is chosen by a queuing system (we chose
Condor for reasons described later), the
software together with all needed libraries is transfered once, so is
the user data. Then the program is executed in a temporary file space
and after the end of the run, the produced data is transfered back to
the user's machine or the the central data server. The red lines shows
the possible institute boundaries in this scenario. As the GRID
protocols gives a uniform access to the resources, the access is the
same no matter in which institute the resource is, but the software
server should be in the same domain as the executing machine to
prevent unnecessary data transfer over WANs.

For us, the main aim of our work is to use the existing toolkits and
the existing collaboration software together and to exploit the full
capacity of all the PCs in our institute. In addition we want to use
this as a testbed to understand the gains and potential problems of
such a GRID. 

The Globus toolkit itself doesn't come with a queuing system, which
is needed to choose a machine to run a software bundle. But several external 
queuing systems are already supported by Globus, for example
LSF, PBS and Condor. Supported means here that Globus knows in
principle how to access these batch systems.

For the choice of the batch system we particularly aimed at the
optimal use of our desktop PCs in our department. This means that

\begin{itemize}
\item during working hours the PCs shouldn't run any jobs when the
user is working,
\item the system should care of PCs, which are switched off for any
reason without any interaction of the system operators,
\item the system should choose a free PC on its own.
\end{itemize}

With our preconditions, the Condor queuing system~\cite{wwwcondor} seems
to be the best choice. The basic Condor system is a
high-throughput-computing
environment whose task is to balance the jobs between machines in a
way that the idle times are minimized. It supports a mechanism, 
which suspends jobs when a running  
machine gets load or interactive work and releases the job again
when idle, so it fits our needs. 
With Condor-G~\cite{wwwcondorg} a GRID enabled
version of Condor is available, but these enhancements are not used
here. The GRID enhancements to Condor were written to track the
users' GRID jobs -- a task which is done here in Wuppertal by our own
system. As Condor is not available as Open-Source software, one would
rely otherwise on the command line tools which may change.

Furthermore, Condor has a file transfer mechanism included, but this
requires that the job
is linked against a special library. This mechanism is not used
here, instead the GRID protocols are used to transfer files. This has
several advantages:

\begin{itemize}
\item using plain Condor requires the binary
executables 
to be linked to the Condor library. This is only possible if one has
at least access to the object files. So one has to distribute two
versions of the software, which can be a disadvantage in a big
collaboration.
\item the Condor traffic has not to be tunneled seperatly through the
firewalls,
\end{itemize}

Although Globus supports Condor as a batch queuing system, some
small modifications had to be applied to the Globus code to get access
to our Condor queue and to optimize the co-operation of these two
software packages.

\subsection{The GRID system in Wuppertal}

Based on the thoughts in the previous chapter, we developed our GRID
in Wuppertal in the following way:

\itemize
\item besides the Globus Toolkit 2, Condor is installed on every
participating PC,
\item we have one machine set up as a central software software. There
are no special requirements to this machine,
\item one machine with Globus installed and appropriate disk space
acts as a central data server. This machine does not need to have
Condor installed, to prevent this important machine to get high load
from batch jobs,
\item one machine out of the normal machine acts as a Condor server. On
this machine, Globus was configured to access the Condor queue(s).

The Condor system itself
can be used by users who don't want or can't use the GRID, but only want
to use the Condor queuing system. Both work smoothly
together. We tested this with an example D\O\  $t\bar{t}$
crossection using the root-Analysis-Framework~\cite{wwwroot}.

Using the GRID, one gains a uniform access to the queuing systems, so
extensions to other institutes (even with queuing systems other than
Condor). That means that from the user side of view, he can submit
jobs with the same command (or within the GUI) without having to care
about how the exact queueing mechanism on the target cluster looks
like. We extended our system to a machine at the Aachen Technical
University and successfully tested the inter-institute
communication. These tests show, that in general modifications to
existing Firewalls are needed. Although the range of used TCP ports
can be limited in the Globus toolkit, it seems that in this stage of
the toolkit all ports above 1024 have to be opened. But implementation
of a virtual private network with a third party software
(i.e. FreeS/WAN~\cite{wwwfreeswan}) seems  however to be unnecessary. 
All communication
between the nodes are encrypted using the OpenSSL
library~\cite{wwwopenssl} and access control is only granted by
presenting a valid and signed X.509 certificate. 
Furthermore, the access to the
systems is always controlled by the local administrator.

\section{The Gridnavigator Program}

This chapter introduces the Gridnavigator software developed in
Wuppertal. The main two goals of this program is to develop software
layers (``sandboxes'') about the existing AMANDA collaboration
software
and to simplify the use of the GRID, which is quite complicated using the
standard commands given by the Globus Toolkit.

The Gridnavigator is very modular and consists of two main 
parts: the already introduced 
sandbox software and a graphical user interface (GUI). The latter
was written to make all the GRID software structure usable to
physicist, which do not have detailed knowledge about the 
GRID terminology, while the sandbox software is a layer around the
standard AMANDA software which takes care about the file transfer,
executes the AMANDA software with all the needed libraries and informs
the user about the job status. The approach of the sandboxes has 
the following advantages:

\begin{itemize}
\item the people developing the standard software can keep their way
of handling in- and output, so that
\item modifications (i.e.~new versions) can be implemented fast and
\item the institutes not participating in the GRID have the usual
software structure.
\item Non-default libraries needed by the AMANDA software are included
  to minimize installation effort on the desktop machines.
\item And - most important - this was much less time consuming than
reprogramming parts of the existing software.
\end{itemize}

These sandboxes are realized so far for the following programs:

\begin{itemize}
\item dCORSIKA~\cite{wwwdcorsika} - an air shower generator (mainly written in FORTRAN)
\item MMC~\cite{wwwmmc} - a code for muon propagation in matter (written in JAVA)
\item AMASIM~\cite{wwwamasim} - a simulation tool for AMANDA (written in C and FORTRAN)
\end{itemize}

The complete program - the sandbox and the GUI - 
is written with the Python programming
language~\cite{python}. As python compiles the program in a byte-code
like JAVA~\cite{java}, the program may be easily ported to other operation
systems. The graphics are produced with the tkinter program library~\cite{wwwtkinter},
which is the standard library for writing graphical user interfaces
in Python and is also available for many platforms.

The JAVA code was compiled using the new
features of the GNU C compiler gcc~\cite{wwwgcc} from version 3
on~\cite{wwwgcc3}. This compiles JAVA into native machine code, which
makes it unnecessary to have a JAVA Runtime Environment (JRE)
installed on every machine. We also implemented successfully a sandbox
with a JRE, but this approach makes the bundles much bigger.

\begin{figure}
  \begin{center}
    \mbox{
            \epsfig{file=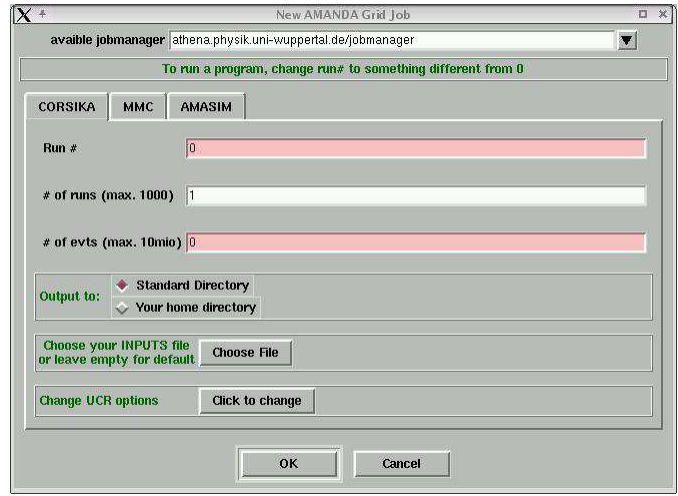,height=6.5cm}
         }
  \end{center}
  \caption[AMANDA standard software dialog]
  {AMANDA standard software dialog}
  \label{amanda}
\end{figure}

Due to the modularity of the Gridnavigator software, the user can
start a software run now via the command line or with the GUI.
Picture~\ref{amanda} shows the input screen of the GUI to run the AMANDA
software. 

The user has the option to get the output back to his home
directory or to a centralized directory at his institute (or to any
other GRID resource he has access to). To get
access to the home directory, a local GASS server\footnote{See
  chapter~\ref{intro} for an explanation of GASS} is started by the
Gridnavigator program. Furthermore, the user
is informed via an email about the start and finish of his run if an
email address was entered. Error messages are also being send this
way. 

All needed parameters like the name and directory of the local data
server, the local mail server to deliver the mails, etc.~are set for
each domain seperatly in the Gridnavigator software, to ensure that
software may run at any resource it has been configured for - and of
course to which the user has access to.

The sandboxes can be used without the GUI, but the GUI itself
uses the sandboxes. Some options of the sandboxes are not
accessible via the GUI, but these options are not widely used. The
separation in the two parts has
the advantage, that in case of heavy
mass production, the jobs can be submitted to the batch queues with a
script or another program. 

\begin{figure}[h]
  \begin{center}
    \mbox{
            \epsfig{file=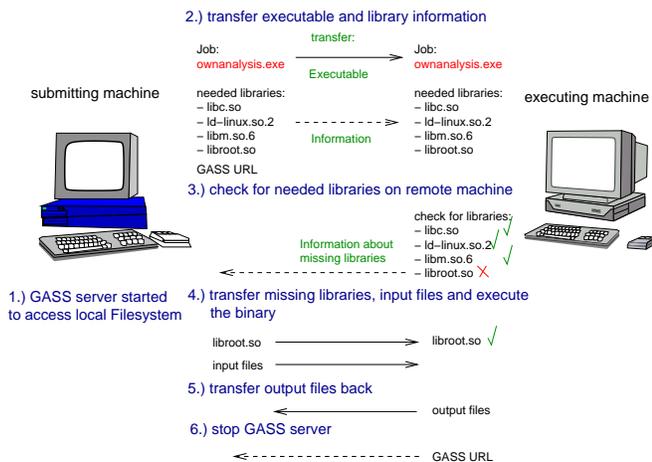,height=6cm}
         }
  \end{center}
  \caption[Scheme of the library resolving mechanism and data transfer mechanism. Special cases not illustrated.]
  {Scheme of the library resolving mechanism and data transfer mechanism. Special cases not illustrated.}
  \label{libraries}
\end{figure}

\begin{figure}
  \begin{center}
    \mbox{
            \epsfig{file=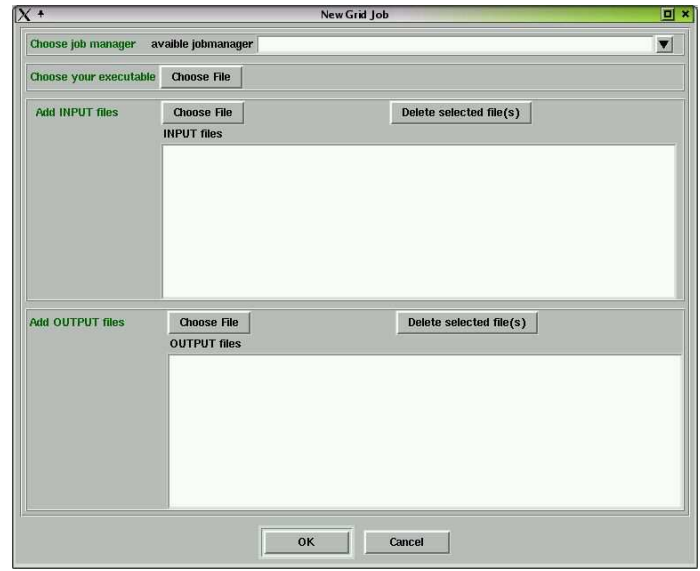,height=7.5cm}
         }
  \end{center}
  \caption[Dialog to run own code in the Grid environment]
  {Dialog to run own code in the Grid environment}
  \label{owncode}
\end{figure}

In addition to the AMANDA standard software, own analysis software 
is also supported, which is in general
not GRID-enabled. For
this reason, another sandbox tool has been written which transfers the in- and
output like it does with the standard software. In this case, the user
has to provide the program name and the names of the in- and output
files. See picture~\ref{owncode} for the input screen. 
The Gridnavigator then takes care about the needed libraries of this
executable. For this reason, a list of all needed libraries is created
on the submitting machine. On the remote host, the sandbox code then
tries to resolve these dependencies and in case not all libraries are
present at this machine, the missing ones are transfered directly from
the submitting machine using the GASS protocol. This even works for
libraries which are not installed system-wide on the submitting
machine. See figure~\ref{libraries} for a scheme of the complete
mechanism. Note that this can be more complex in special cases, where a
third machine is involved.

This mechanism is in particular a big improvement if one works accross
boundaries of just one 
institute, as it makes it unnecessary to install every
library needed by the collaboration (maybe even only by one member of
the collaboration) at every institute and make the
GRID usable even in the case where a user wrote his own code using
libraries not installed anywhere else than on his local
machine. Furthermore, this mechanism can help keeping the software
repository up-to-date, as new versions of the collaboration software
can be integrated very easy. We chose to bundle the libraries of the
standard software for now to make the system more fault tolerant.

For his own code, the user has specified the information about in- and output
files once, but then 
he gains the advantage of
having his code running even at remote side(s) without any further
modifications.

\begin{figure}
  \begin{center}
    \mbox{
            \epsfig{file=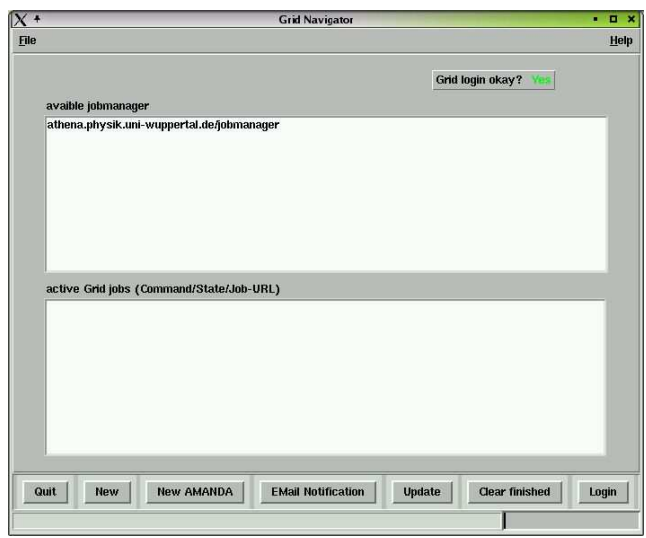,height=7cm}
         }
  \end{center}
  \caption[The main screen of the Gridnavigator program]
  {The main screen of the Gridnavigator program}
  \label{mainscreen}
\end{figure}

Besides this, one more part is covered by the Gridnavigator
GUI:

It tracks all the submitted jobs and informs the user about the
status of his jobs. See picture~\ref{mainscreen} for this. 
As the Globus toolkit can access different queuing systems, it
cannot provide simply a list of all running jobs. Instead it provides
the user with a URL\footnote{URL = Uniform Resource Locator. The URL
have the same format as widely used in the worldwide web (WWW)}, with
can be used to resolve the status of the job. Therefore, a list of these
job-URLs is stored together with other information by the
Gridnavigator program. The user may cancel one or more
jobs.

The program also takes care about the creation of the necessary proxy,
this is a time-limited cryptographic certificate which enables the user
to access the Grid resources. See~\cite{proxy} for details of this
mechanism. The user can ``log-in'' to the Grid using the GUI and may
specify the lifetime of the proxy.

\section{Experience and Advantages}

We installed this system on 18 PCs in our institute. These have
several versions of the LINUX Operating Systems (SuSE Linux Versions
7.3, 8.0 and 8.1 and RedHat 7.x) and different
configurations concerning disk space, memory and CPU. No general
problems occurred. 

One problem occurred when trying to run the AMAsim software
package. This package requires quite a lot amount of memory
(approx.~512 MB per instance). When such a job was started on a PC
which less RAM, then this disturbs the normal operation quite a
lot, due to the swapping activity of the LINUX operating system. We
solved this situation by defining two Condor queues, one consists of
all PCs, one only included the well-equipped ones. On the Condor
server, another jobmanager was defined with special parameters passed
to the Condor queuing system. A jobmanager is a gateway between the
GRID and a queuing system in the Globus Toolkit.
Sending AMAsim jobs to the one queue with the special parameters and
all others to the general queue solved now our problem.

With this system, we achieved in average 
200 million primary CORSIKA events per week with one mouse click. The
situation before was running CORSIKA on several Sun UltraSPARC II
workstations, where we could get around 15 million primary events per
week with a huge overload in administration (connect to every single
machine, start the program, transfer back the output, etc.).

We extended the system to the Technical University of Aachen, where we
had thanks to Dr.~Rolf Steinberg the possibility to test
inter-institute communication. This succeeded when opened all TCP
ports 
above 1024 on the
institutes' firewalls. The Certification
Authority could be installed, so that users from one side had
automatically access to the GRID resources on the other side. We could
successfully test our software layers (``sandboxes''), so AMANDA
software could run in Aachen, although it was not seperatly
installed. This system served 4 AMANDA diploma students in Wuppertal as
basis for their Monte Carlo production. 

The example D\O\  analysis using the plain Condor environment 
archived 2000 grid points
per hour. 

We tested successfully the library resolving mechanism using some
software of the
DELPHI collaboration. DELPHI is one of the former experiments operated
at the LEP electron-positron accellarator at CERN, Geneva.

\section{Summary}

We presented a way how an existing GRID toolkit - the Globus 
Toolkit 2 in our case - can be used to
establish and run a GRID system within a physics collaboration and how the
physicists can profit from these software structures.

We established successfully
such a system in Wuppertal using the Globus toolkit and the Condor
queuing system. We presented our approach to build up a GRID with a
software server and a data storage server.

We extended the system with success to other institutes.

We showed an approach of how to run existing collaboration software,
which is not GRID-enabled yet, in such a GRID environment. We
presented the ``Gridnavigator'' software, which was developed in
Wuppertal. This software has two parts: some ``sandboxes'' and a
graphical user interface.

The ``sandboxes'' are software layers around the existing collaboration
software, which takes care about the needed libraries, the file
transfer, the user notification, etc. A similar software was written to
make user code usable, this is code which has not been bundled by
anybody to run in a GRID environment and which is not
GRID-enabled. This sandbox cares - like in the previous case - about
the file transfer, the execution in a temporary file space and the user
notification. Furthermore, it resolves library dependencies by
transferring those libraries which are not installed on the target
system from the machine where the job was submitted from. This is in
most cases the machine where the software was build on. This feature
is very important nowadays, where most of the existing software has not
been rewritten to be used in a GRID environment.

For an easy access to the system,
we presented the Gridnavigator program, a graphical user interface to
our Grid system. Using that, the user can submit bundled and own
software without having to care about the underlying GRID software
structure. Furthermore, it keeps track of submitted jobs and gives the
user the possibility to cancel one or more of his jobs.

We tested the ``Gridnavigator'' with the three main parts of the
AMANDA offline software chain. These work smoothly within the GRID
environment. Furthermore, we tested the support of own user software
by running code from the DELPHI collaboration within the GRID. And we
tested that the Condor queuing system itself can be used without
using the GRID by running a D\O\  analysis run.

Our work now gives the AMANDA collaboration the possibility to build
up a ``virtual organization'' by connecting the participating
institutes with the Globus toolkit and use our program to access the
GRID without knowledge of the underlying GRID software layer. Jobs can
be submitted to foreign institutes when local resources do not fit
the needs of the user. Access to produced mass production data can be
unified.

\bibliographystyle{own_unsrt}
\bibliography{papier}

\begin{thebibliography}{10}

\bibitem{wwwglobus}
http://www.globus.org.

\bibitem{anatomy}
I.~Foster, C.~Kesselman, and S.~Tuecke.
\newblock The Anatomy of the Grid: Enabling Scalable Virtual Organizations.
\newblock International J. Supercomputer Applications, 15(3), 2001.

\bibitem{wwwsam}
Andrew Baranovski, Diana Bonham, Gabriele Garzoglio, Chris Jozwiak,
  Lauri~Loebel Carpenter, Lee Lueking, Carmenita Moore, Ruth Pordes, Heidi
  Schellman, Igor Terekhov, Matthew Vranicar, Sinisa Veseli, Stephen White, and
  Victoria White.
\newblock SAM Managed Cache and Processing for Clusters in a Worldwide
  Grid-Enabled System, Proceedings of the Large Cluster Computing Workshop
  2002.

\bibitem{foster}
Ian Foster and Carl Kesselman.
\newblock The Grid: Blueprint for a New Computing Infrastructure.
\newblock Morgan Kaufmann Publishers, July 1998.

\bibitem{SECURITY}
I.~Foster, C.~Kesselman, G.~Tsudik, and S.~Tuecke.
\newblock A security architecture for computational grids.
\newblock in ACM Conference on Computers and Security, pages 83-91. ACM Press,
  1998.

\bibitem{GASS}
J.~Bester, I.~Foster, C.~Kesselman, J.~Tedesco, and S.~Tuecke.
\newblock GASS: A Data Movement and Access Service for Wide Area Computing
  Systems.
\newblock Sixth Workshop on I/O in Parallel and Distributed Systems, May 1999.

\bibitem{gridftp}
W.~Allcock, J.~Bester, J.~Bresnahan, A.~Chervenak, L.~Liming, S.~Meder, and
  S.~Tuecke.
\newblock GGF GridFTP Working Group Document, September 2002.

\bibitem{MDS}
S.~Fitzgerald, I.~Foster, C.~Kesselman, G.~von Laszewski, W.~Smith, and
  S.~Tuecke.
\newblock A Directory Service for Configuring High-Performance Distributed
  Computations.
\newblock Sixth IEEE Symp. on High-Performance Distributed Computing, 1997.

\bibitem{HBM}
P.~Stelling, I.~Foster, C.~Kesselman, C.Lee, and G.~von Laszewski.
\newblock A Fault Detection Service for Wide Area Distributed Computations.
\newblock Proc. 7th IEEE Symp. on High Performance Distributed Computing, pages
  268-278, 1998.

\bibitem{wwwcondor}
http://www.cs.wisc.edu/condor.

\bibitem{wwwcondorg}
http://www.cs.wisc.edu/condor/condorg.

\bibitem{wwwroot}
http://root.cern.ch.

\bibitem{wwwfreeswan}
http://www.freeswan.org.

\bibitem{wwwopenssl}
http://www.openssl.org.

\bibitem{wwwdcorsika}
http://amanda.uni-wuppertal.de/\~{}dima/work/CORSIKA.

\bibitem{wwwmmc}
http://amanda.uni-wuppertal.de/\~{}dima/work/MUONPR.

\bibitem{wwwamasim}
http://www.ps.uci.edu/\~{}hundert/aman\-da/amasim/amasim.html.

\bibitem{python}
http://www.python.org.

\bibitem{java}
http://java.sun.com.

\bibitem{wwwtkinter}
http://www.python.org/topics/tkinter/.

\bibitem{wwwgcc}
http://gcc.gnu.org.

\bibitem{wwwgcc3}
http://gcc.gnu.org/gcc-3.0/features.html.

\bibitem{proxy}
http://www.globus.org/Security.

\end{thebibliography}





\end{document}